\documentclass[sigconf]{acmart}

\AtBeginDocument{%
  \providecommand\BibTeX{{%
    \normalfont B\kern-0.5em{\scshape i\kern-0.25em b}\kern-0.8em\TeX}}}

\setcopyright{none}
\acmConference[USRW Workshop at RecSys'26]{Unified Search and Recommendation Workshop}{October 2, 2026}{Minneapolis, MN, USA}
\renewcommand\footnotetextcopyrightpermission[1]{} 

\begin{document}

\title{Every Article Deserves a Video: Contextual Video Matching for Digital Publishers}

\author{Arnaud Corone}
\email{arnaud.corone@dailymotion.com}
\affiliation{%
  \institution{Dailymotion}
  \city{Paris}
  \country{France}
}

\author{Brice Pierre de la Brière}
\email{brice.delabriere@dailymotion.com}
\affiliation{%
  \institution{Dailymotion}
  \city{Paris}
  \country{France}
}

\author{Gladys Roch}
\email{gladys.roch@dailymotion.com}
\affiliation{%
  \institution{Dailymotion}
  \city{Paris}
  \country{France}
}

\author{Samuel Leonardo Gracio}
\email{samuel.leonardogracio@dailymotion.com}
\affiliation{%
  \institution{Dailymotion}
  \city{Paris}
  \country{France}
}

\author{Yassine Bouher}
\email{yassine.bouher@dailymotion.com}
\affiliation{%
  \institution{Dailymotion}
  \city{Paris}
  \country{France}
}

\author{Parvati Chauchaix}
\email{parvati.chauchaix@dailymotion.com}
\affiliation{%
  \institution{Dailymotion}
  \city{Paris}
  \country{France}
}
\renewcommand{\shortauthors}{Arnaud Corone, Brice Pierre de la Brière et al.}

\begin{abstract}
As digital publishers face the challenge of managing massive content catalogs, the ability to effectively embed relevant video within text-based articles has become critical for both monetization and user retention. However, manual selection is impractical for large-scale publishers, especially when navigating their own extensive video libraries or the entire global Dailymotion catalog. In this paper, we present the ``Contextual Video Matching'' system, a solution that automatically matches relevant videos with text-heavy web pages and articles. By leveraging Large Language Models (LLMs) and textual embeddings, we provide a scalable solution for publishers to efficiently combine video content with their articles. We discuss in detail the motivations, architecture, evaluations, and deployment of this system within Dailymotion's production environment. Since its launch, the system has been adopted by hundreds of publishers, significantly increasing user engagement and enriching user experiences with highly relevant video content.
\end{abstract}

\maketitle

\section{Introduction}
Dailymotion is a global video streaming service and platform that connects hundreds of millions of active users to a vast catalog of hundreds of millions of videos, generating billions of views every month around the world. Beyond its consumer-facing platform, Dailymotion provides a video player technology that allows publishers to host, distribute, and monetize their content directly within their own domains.

\begin{figure}[t]
    \centering
    \includegraphics[width=0.8\columnwidth]{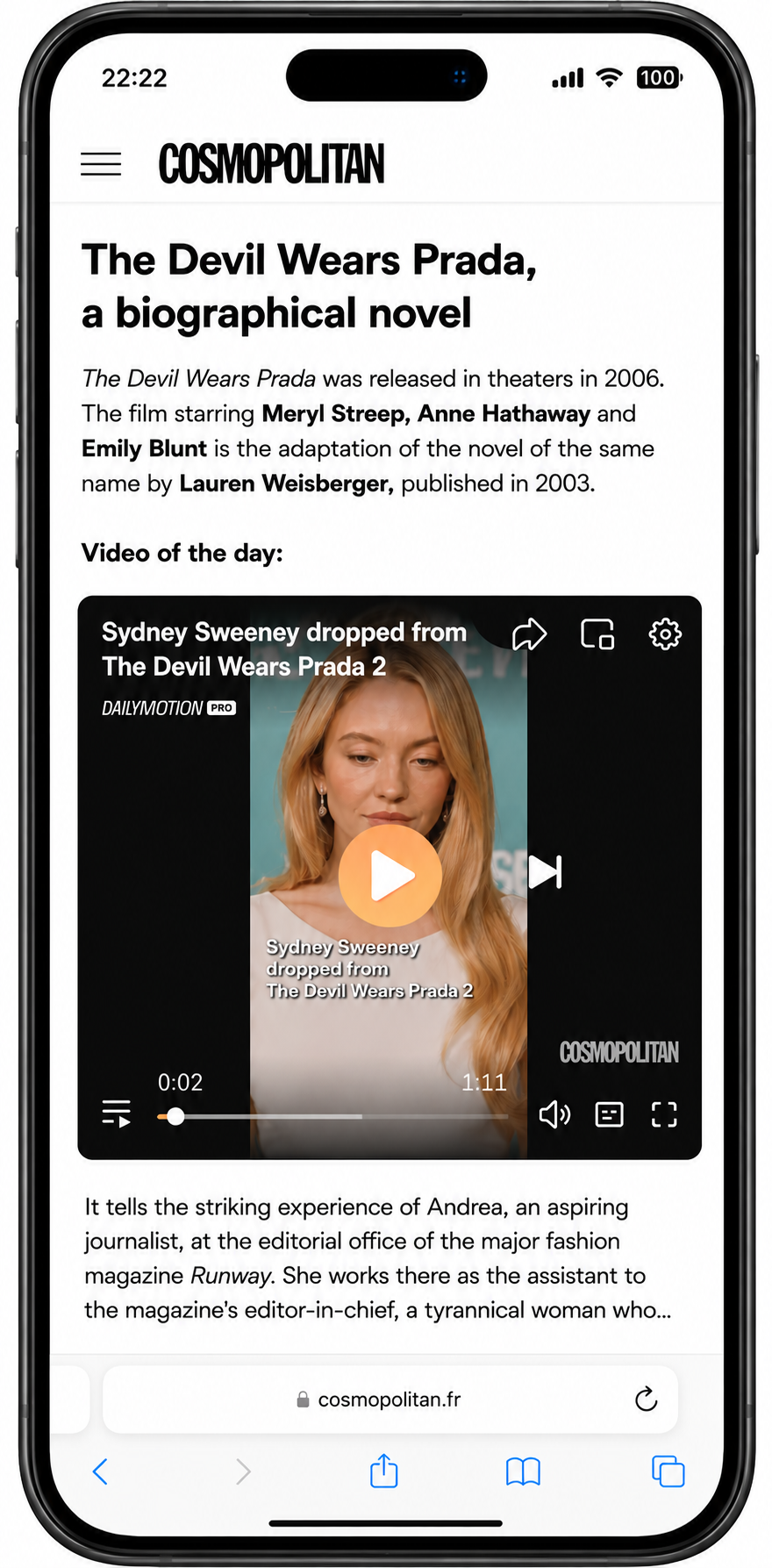}
    \caption{Real-world production example of our Contextual Video Matching system integrated on a publisher's page. (text translated from French for readability)}
    \label{fig:publisher_example}
\end{figure}

Digital publishers are increasingly adopting a multimodal approach by integrating video into articles. This shift is driven by three primary factors: (i) enhanced monetization, as in-stream video CPMs significantly outperform traditional display advertising~\cite{iab2026report}; (ii) SEO optimization via increased dwell time; and (iii) maximized ROI through the repurposing of social media assets into a continuous media ecosystem on proprietary platforms. However, relying on editorial teams to manually pair videos with every newly published article creates an operational bottleneck that cannot scale. Although dedicated videos are straightforward to embed, identifying relevant content from millions of potential matches or discovering hidden gems within a publisher’s own archive requires automated contextual intelligence to remain operationally scalable. Addressing this challenge represents a growth opportunity: by deploying a matching solution, we can enable video integration on pages that would otherwise remain text-only. This creates a mutually beneficial model that generates incremental revenue and engagement for both the publisher and the Dailymotion platform. Figure~\ref{fig:publisher_example} illustrates an example of our system deployed in production, matching a contextually relevant video to a publisher's article.

Scaling video integration requires matching full-length articles with relevant video. While there exists a diverse literature on Video-Text Matching~\cite{zhu2023videotext}, most of the research focuses on aligning videos with short text snippets, such as single sentences or search queries~\cite{radford2021clip, xu2021videoclip, bain2021frozen} rather than long-form text. To fill this gap, we introduce the Contextual Video Matching, an end-to-end system designed to pair full-length articles with relevant videos at scale. This task is still not explicitly studied in the existing literature to our knowledge, neither in information retrieval~\cite{mitra2018neuralir} nor in the recommendation fields~\cite{lops2011content}. This system has shown great adoption and satisfaction from our publishers.

The primary contributions of this work are as follows:
\begin{itemize}
    \item \textbf{System Design:} We introduce a scalable architecture that leverages Large Language Models (LLMs) and vector embeddings to bridge the semantic gap between long-form text and video.
    \item \textbf{Industrial Deployment:} We detail the challenges of deploying this system within a high-traffic production environment supporting hundreds of publishers.
    \item \textbf{Impact Analysis:} We demonstrate through online A/B testing how our system improves page quality, increases user engagement, and enriches the overall user experience.
\end{itemize}

The remainder of this paper describes the Contextual Video Matching architecture, its deployment, and the results.


\begin{figure*}[t]
  \centering
  \includegraphics[width=\textwidth]{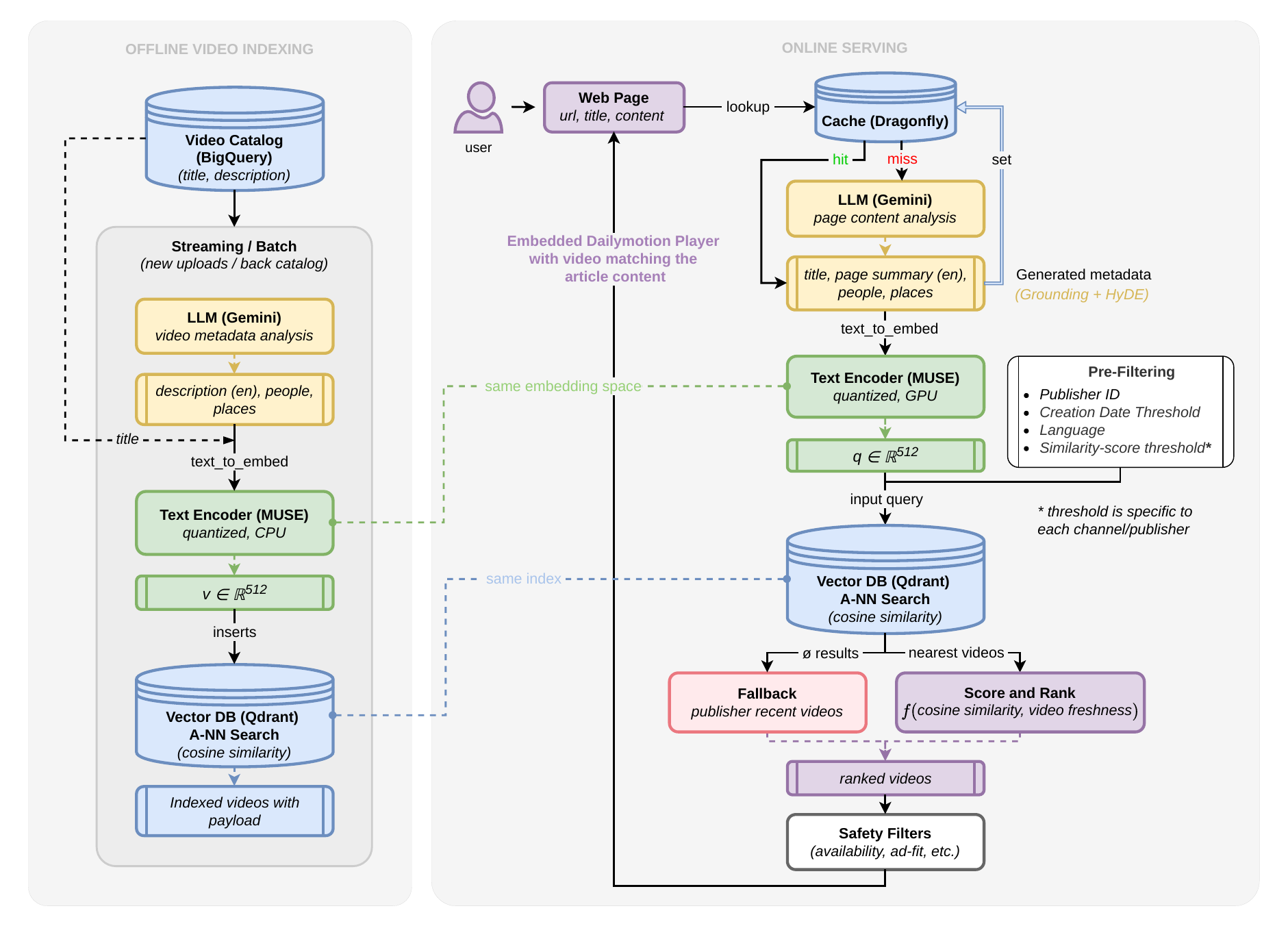}
  \caption{Contextual Video Matching system architecture. The offline pipeline (left) processes the video catalog, while the online serving pipeline (right) handles real-time requests for webpage and video matching.}
  \label{fig:p2v_schema}
\end{figure*}

\section{Contextual Video Matching: System Design}
This section describes the Contextual Video Matching system, an end-to-end architecture designed to automate contextual media selection at scale.

\subsection{The Need for Relevant Videos}
Publishers often possess vast libraries of video content alongside a large volume of articles. Manually matching videos with articles introduces severe operational friction. In many cases, the author of the page may not even be fully aware of the extent of their own video inventory. Furthermore, irrelevant video placement would negatively impact user experience and brand integrity. Consequently, maintaining strict contextual alignment is the primary driver of successful product adoption and publisher trust. The recommended videos must be highly pertinent to the specific article's content. Our system addresses this bottleneck and strict relevancy constraint, thus improving the page quality and unlocking new revenue streams.

\subsection{How to Match Articles with Videos}
The overall architecture of the Contextual Video Matching system is structured as a multi-stage pipeline, as illustrated in Figure~\ref{fig:p2v_schema}. The pipeline is designed to process web pages in real-time through three core stages: structural text extraction, semantic synthesis, and vector-based candidate retrieval.

\subsubsection{Extracting and formatting information from webpages}
\label{parsing}
Since the Contextual Video Matching system must process articles from a wide variety of publisher websites, the first challenge is to extract the semantic essence of a page while filtering out noisy HTML elements like advertisements, navigation menus, and sidebars. To do so, we feed the raw HTML body of the page to an LLM (specifically, gemini-2.5-flash) whose task is to distinguish between meaningful and irrelevant parts in order to summarize the article. Importantly, the output of the LLM is framed to be in the format of a video metadata, consisting of a title and a concise description. Inspired by the Hypothetical Document Embeddings (HyDE) framework described by Gao et al. in~\cite{gao-etal-2023-precise}, we implement a synthesis layer where the LLM is prompted to generate hypothetical video metadata that would perfectly align with the article. This redefines the retrieval task as a symmetrical matching process between a hypothetical perfect video and the dataset of real videos available.

\subsubsection{Enhancing and grounding article information}
\label{grounding}
The global nature of our publisher network and video inventory requires a multi-lingual embedding model to maintain high-quality matching across dozens of languages. However, while multilingual embeddings can project articles and videos from different languages into the same shared space, for relevance purposes, we must ensure that the video and article languages match. Consequently, the LLM is prompted to explicitly extract the article's language which is subsequently used to filter candidates during retrieval.

For text representation at Dailymotion, we rely on the Multilingual Universal Sentence Encoder (MUSE) \cite{yang2020muse, cer2018use} across nearly all our textual embedding tasks. Despite its 2019 release, MUSE has consistently outperformed more recent models on our specific datasets. Beyond its performance, MUSE offers critical industrial advantages: it is open-source, straightforward to deploy, and maintains a low latency footprint during real-time serving. Furthermore, MUSE currently serves as the backbone of our video indexing pipeline, having been used to embed our entire catalog of 200 million videos. This represents a significant pre-existing infrastructure constraint; re-encoding such a vast library with a newer model for every experimental iteration would be computationally expensive and operationally complex. Consequently, our research focuses on optimizing the input strategies to maximize the retrieval quality within this established embedding space. To achieve this, we augment the HyDE framework described in Section~\ref{parsing} to also overcome the limitations of the embedding model's age.

Since MUSE was frozen in 2019, it lacks awareness of entities and events that emerged after its training. As articles may deal with possibly very recent context, we also leverage the grounding capabilities of recent LLMs to define the most important persons, places or events mentioned in the page within the synthesis layer. For instance, the LLM can describe a recently elected politician or a newly created sports league in a few words, providing contextual information that our embedding model, trained before these entities emerged, would otherwise lack. This approach is related to recent work on emerging entity representation: Ghonim et al. \cite{ghonim2025raed} generate descriptions of entities absent from static knowledge bases using retrieval-augmented LLMs, showing that such textual grounding helps downstream tasks even without retraining the underlying model. More broadly, a growing body of work explores using LLMs to augment content representations for recommendation. Xi et al. \cite{xi2024kar} leverage LLMs to generate factual and reasoning knowledge that is encoded into augmented vectors for recommendation, while Liu et al. \cite{liu2024once} use LLMs to summarize and enrich item descriptions for content-based recommendation. Other techniques improve the query or the document at retrieval time \cite{lewis2020rag}. Our approach is specifically designed to bridge the temporal knowledge gap of a deployed embedding model. So, rather than enriching features for a trainable recommender, we use the LLM's up-to-date world knowledge to produce short grounding descriptions that make recent entities interpretable by an older, frozen encoder, without the need for retraining.

\subsubsection{Embedding videos and pages}
Once the article's hypothetical video metadata consisting of the title, description, and grounded entities has been generated, it is mapped into the embedding space using the MUSE model. This step ensures that the representation of the article is perfectly aligned with the existing video catalog. On the retrieval side, the actual video embeddings are persistently stored in a vector database. Then, we compute the cosine similarity score between the article’s hypothetical embedding and the video catalog to retrieve the approximate nearest neighbors~\cite{karpukhin2020dpr, malkov2020hnsw}. This allows us to efficiently surface the videos that share the highest semantic overlap with our article.

\subsubsection{Adapting the results for each publisher}
\begin{figure}[h]
    \centering
    \includegraphics[width=\columnwidth]{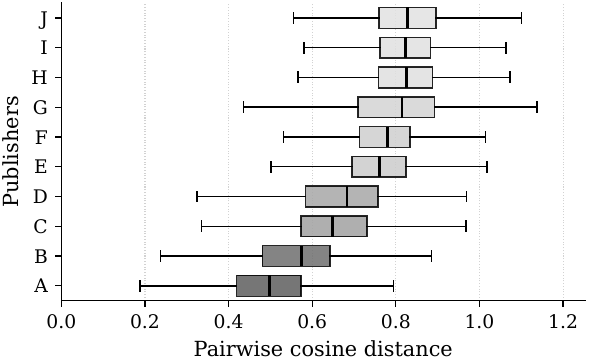}
    \caption{Semantic diversity across the top 10 publishers (A--J). Distances are computed via pairwise cosine similarity on MUSE embeddings of the 200 most recent videos.}
    \label{fig:diversity}
\end{figure}

To ensure high precision, we apply a threshold to the cosine similarity between article and video embeddings. This threshold is dynamically defined for each publisher as a single global threshold cannot effectively serve all. As illustrated in Figure~\ref{fig:diversity}, the distribution of pairwise cosine distances varies significantly across publishers. For instance, a niche publisher specialized in a single topic such as sports will have a much denser embedding space than a generalist news outlet covering a vast array of subjects. In a dense space, a high similarity score might still yield several mediocre matches, whereas in a sparse one, a lower score could retrieve a highly relevant outlier. To account for this, we dynamically compute thresholds based on each publisher's specific embedding spread, using the 75-th percentile of the distribution. Finally, publishers can manually override these thresholds to align the system with their specific editorial standards and relevance tolerance.

Furthermore, for most publishers, aspects other than relevancy come into consideration, such as freshness of the video or its performance. Hence, taking inspiration from recommender methods \cite{covington2016youtube, yi2019twotower}, we customized a score mixing the cosine similarity between the embedding of the article and the videos with other metrics to rank the nearest neighbors passing the threshold, allowing us to select the best video to display on the page.

\section{Offline Evaluation}
This section details our evaluation framework, from the definition of our production dataset to the automated strategies developed to measure semantic relevance.
\subsection{Dataset and Evaluation Challenges}

\subsubsection{Dataset Description.} Our evaluation relies on a diverse production dataset of 1,032 web pages sourced from Dailymotion’s global publisher network. This corpus covers a broad spectrum of thematic domains---such as news, sports, and entertainment---thereby reflecting real-world distribution.

\subsubsection{Absence of Ground Truth} A significant hurdle in this industrial application is the lack of pre-existing labeled data. To the best of our knowledge, existing public datasets primarily focus on short-form captions rather than long-form editorial content, making them unsuitable for our use case. Furthermore, manual labeling is impractical due to constraints of both scale and subjectivity. On one hand, the volume of daily articles and videos makes manual cross-referencing impossible to sustain over time. On the other hand, defining relevance requires expert editorial synthesis rather than simple keyword matching, making human consensus difficult to achieve at scale. To address these constraints, we implement an automated evaluation proxy by leveraging LLMs-As-A-Judge to assess recommendation quality at scale.

\subsection{LLMs-As-A-Judge}
We employed Gemini 3.1 Pro as our primary evaluator, ensuring that the judge model possesses higher reasoning capabilities than the models being evaluated~\cite{zheng2023llmjudge}.
Our framework relies on two complementary evaluation protocols.
\subsubsection{Pointwise Multidimensional Scoring}
In this protocol, the LLM-Judge evaluates the relevance of individual video recommendations based on three core principles:
\begin{itemize}
    \item \textbf{Granular Scoring:} We prompt the judge to map each article-video pair onto a 1--5 Likert scale. The scoring logic is defined as follows: a score of 1 denotes entirely unrelated content (e.g., financial news paired with a cooking tutorial), while a 3 indicates a shared general topic with no situational link (e.g., two unrelated space exploration reports). Finally, a score of 5 is reserved for direct editorial continuations (e.g., a sports game report paired with the winner's post-game interview).

    \item \textbf{Multi-dimensional Alignment:} To prevent the judge from focusing solely on superficial keyword matching, the prompt decomposes relevance into two distinct axes: \textit{(A) Topical Relevance} (shared events and entities) and \textit{(B) Reader Intent} (the logical utility of the video for the reader), as recommended by Liu et al~\cite{liu-etal-2023-g}. This structured rubric ensures that the system distinguishes between a generic domain match (e.g., a generic sports video paired with a football article) and a high-precision editorial match.
    \item \textbf{Reasoning-before-Rating (CoT):} We enforce a Chain-of-Thought reasoning workflow~\cite{wei2022chain} to ensure the judge justifies its alignment decision before producing the score.
\end{itemize}

\subsubsection{Pairwise Comparative Evaluation}
To determine the relative effectiveness of our recommendation strategies, we implemented a tournament style framework based on the Bradley-Terry model~\cite{bradley1952rank}. 

\begin{itemize}
    \item \textbf{Comparison Protocol:} For a given article, the LLM-Judge is presented with two candidate videos from different matching methods. It must decide which is more relevant or declare a tie.
    \item \textbf{Elo-like Ratings}: We use Maximum Likelihood Estimation (MLE) to convert these win/loss/tie outcomes into global scores (centered at 1500), providing a clear ranking of the compared strategies.
    \item \textbf{Statistical Robustness:} To ensure statistical rigor and account for the inherent correlation of pairwise comparisons within the same article, we compute 95\% Confidence Intervals using a cluster bootstrap protocol~\cite{cameron2008bootstrap}. By resampling at the article level rather than the duel level, we preserve the dependency between comparisons of the same article, yielding a realistic estimation of variance for our final scores.
    \item \textbf{Bias Control:} To prevent position bias, the tendency of LLMs to favor the first option~\cite{zheng2023llmjudge}, the presentation order of candidates is randomized for every duel.
\end{itemize}

\subsection{Experimental Results}
\label{baselines}
To validate the relevance of Contextual Video Matching before deployment, we benchmarked our system against six increasingly sophisticated baselines:
\begin{itemize}
    \item \textbf{Random:} A video is selected at random from the publisher’s library.
    \item \textbf{Most Recent:} A heuristic-based baseline that selects the latest video uploaded by the publisher.
    \item \textbf{Raw HTML:} MUSE embeddings are generated directly from the article's raw HTML body.
    \item \textbf{Basic HTML Parsing:} This baseline extracts the visible raw text from an HTML page by removing invisible code like scripts and styles.
    \item \textbf{Basic Summary:} An LLM generates a basic summary of the page which is then encoded by MUSE for retrieval.
    \item \textbf{HyDE}: Applies the HyDE framework to generate hypothetical video metadata.
    \item \textbf{HyDE + Grounded Parsing (Ours):} Our proposed approach, which leverages the HyDE framework and grounded LLM parsing to refine the article's semantic representation and optimize retrieval, which is then encoded by MUSE and used for retrieval.
\end{itemize}

\begin{table}[h]
\centering
\caption{Pointwise Evaluation Results: Comparison of Contextual Video Matching Variants and Baselines (1--5 Likert Scale).}
\label{tab:pointwise_results}
\footnotesize 
\begin{tabular}{lcccc} 
\hline
\textbf{Method} & \textbf{Topical} & \textbf{Intent} & \textbf{Mean} & \textbf{Score $\ge$ 4} \\ \hline
Random & 1.124 & 1.079 & 1.121 & 0.0\% \\
Most Recent & 1.285 & 1.227 & 1.280 & 1.1\% \\ \hline
Raw HTML & 1.404 & 1.338 & 1.402 & 2.1\% \\
Basic HTML Parsing & 2.318 & 2.172 & 2.303 & 15.5\% \\
Basic Summary & 2.380 & 2.247 & 2.364 & 16.3\% \\
HyDE & 2.469 & 2.340 & 2.444 & 19.7\% \\
\textbf{HyDE + Grounded Parsing (Ours)} & \textbf{2.535} & \textbf{2.419} & \textbf{2.509} & \textbf{19.9\%} \\ \hline
\end{tabular}
\end{table}

\begin{figure}[h] 
    \centering
    \includegraphics[width=\columnwidth]{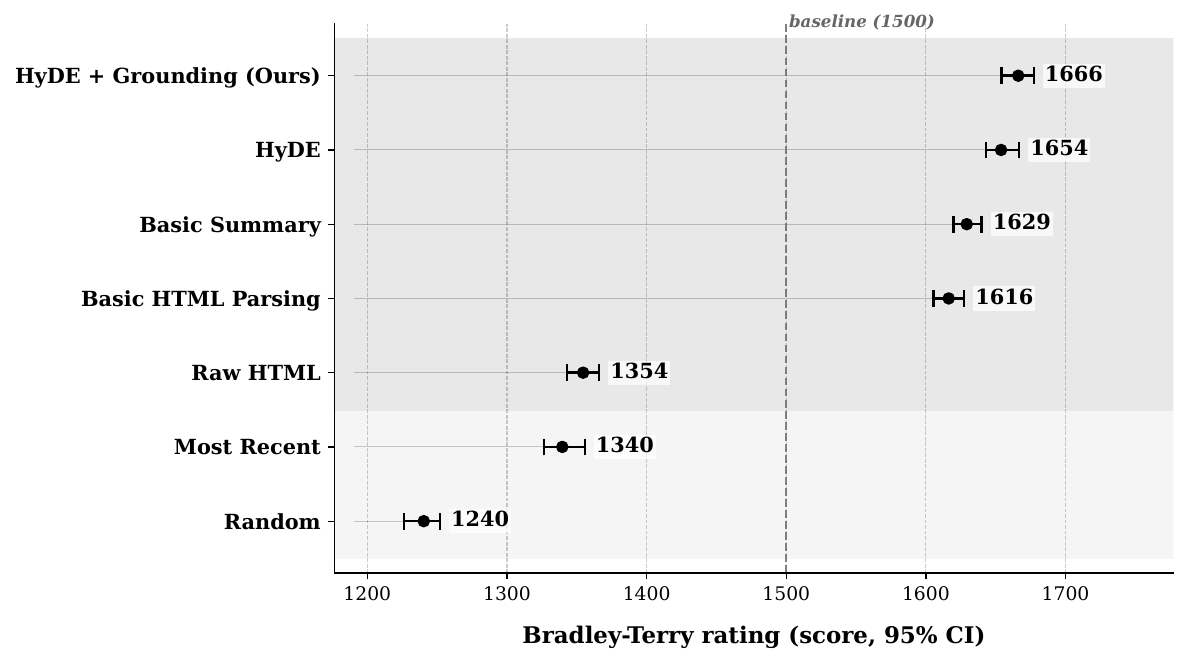}
    \caption{Bradley-Terry ratings for Contextual Video Matching and baselines. Error bars indicate 95\% confidence intervals (cluster bootstrap, 1000 iterations)}
    \label{fig:elo_results}
\end{figure}

As shown in Table~\ref{tab:pointwise_results}, our Contextual Video Matching consistently outperforms all baselines. Although heuristic methods such as Random and Most Recent fail to provide relevant matches (Mean $\approx$ 1.1–1.3), the introduction of HyDE and Grounded Parsing provides a major lift, reaching a Mean score of 2.51. In particular, Contextual Video Matching provides a substantial boost to the percentage of highly relevant matches (score $\ge$ 4) compared to the basic HTML parsing baseline, jumping from 15.5\% to 19.9\%. The Bradley-Terry ratings (Figure~\ref{fig:elo_results}) confirm this superiority. Contextual Video Matching achieves the highest rating (1666), significantly outdistancing the production baselines. The tight 95\% confidence intervals, calculated using cluster bootstrap, demonstrate that these gains are statistically robust across our diverse dataset of long-form editorial articles. 

However, the absolute mean scores from the pointwise relevancy scoring remain moderate, highlighting that even the best-performing model frequently encounters low-relevance candidates. This observation confirms the necessity of our dynamic thresholding strategy (Section 2.2.4) to filter results before they are served to publishers.

\section{Online Experiments}
To demonstrate how video relevancy directly enhances user experience and increases user engagement on the publisher's page, we conducted an online A/B test within Dailymotion’s publisher network. The experiment ran continuously for 10 days on a controlled subset of production traffic. Eligible visitors were randomly and persistently assigned at the user level to either a control group or a treatment group. The control group was served using our random baseline described in Section~\ref{baselines}, representing the least relevant approach, while the treatment group used our Contextual Video Matching, designed to maximize semantic alignment. The experiment involved tens of millions of visitors in each group and tens of thousands of web pages across multiple publishers. To isolate qualified user intent, our analysis focused strictly on non-autoplay environments where users must manually initiate video playback. To quantify this behavioral impact, we analyzed two key metrics:

\begin{itemize}
    \item \textbf{Watch Time per Visitor:} Measures the total time a unique visitor spends watching videos. Under our system, this metric achieved \textbf{+19\% relative increase} compared to the random baseline showing that users stay engaged longer on the page.
    \item \textbf{Average Watch Time per View:} Measures the average duration of a single video playback. This metric demonstrated a substantial \textbf{+21\% relative increase} when powered by our solution, proving that the recommended content is highly relevant to the user.
\end{itemize}

The A/B test results strongly validate our offline evaluation, proving that shifting from random selection to semantic alignment drastically improves the quality of content consumption, while successfully sustaining user attention and increasing the overall time spent on the publisher’s page.

\section{Deployment Challenges and Real-World Impact}
This section describes the practical considerations for deploying Contextual Video Matching at scale. Transitioning from a research prototype to a production system serving hundreds of publishers required addressing significant constraints regarding latency, cost-efficiency, and content extraction.

\subsection{Implementation Choices and Infrastructure}
To serve Contextual Video Matching while maintaining a premium user experience and a sustainable cost structure, we implemented the following strategies:

\subsubsection{LLM-based Content Extraction:} By leveraging the LLM-based parsing detailed in Section~\ref{parsing}, we eliminate the need for domain-specific scrapers. This unified approach allows for instant onboarding of new publishers, as the system automatically adapts to any HTML structure without manual configuration or maintenance overhead.

\subsubsection{Smart Caching and Latency Mitigation} Generating an LLM-based synthesis for every page view is prohibitive due to high inference costs driven by the large token count of raw HTML and the latency requirements of real-time web delivery. Since editorial content is mostly static, we implement a caching layer for the LLM's output, which serves as a hypothetical video proxy. Only the first request—typically triggered by the publisher's editor upon publication—incurs the LLM's processing time. All subsequent users are served from the cache, bypassing the LLM computation step. This architecture provides three key advantages:
\begin{itemize}
    \item \textbf{Minimal Latency:}  Serving most requests without having to wait for the LLM generation, which is a rather slow process, significantly improves the response latency of our system.
    \item \textbf{Cost Efficiency:} LLM calls can be a significant part of the costs associated with running a system like the Contextual Video Matching. With the caching mechanism, whether a page receives ten or ten million visits, the LLM is only billed once. Decoupling LLM calls from traffic volume drastically reduces operational overhead.
    \item \textbf{Dynamic Retrieval:} By caching the hypothetical document representation rather than the final video item, the system remains dynamic. It can recommend newly uploaded, more relevant videos as they become available in the catalog without re-invoking the LLM.
\end{itemize}

\subsubsection{Vector Search Infrastructure}
For the retrieval phase, given the size of Dailymotion catalog, we use a vectorial database optimized for approximate nearest neighbor search. Several options exist for that and were considered \cite{douze2024faiss}. We have opted for Qdrant \cite{qdrant2021}, an open source solution that we already use in different projects and that we can deploy ourselves on Kubernetes clusters.

\subsection{Contextual Video Matching adoption}
Since its global rollout, Contextual Video Matching has seen rapid adoption across Dailymotion’s ecosystem. The system is currently leveraged by over 200 publishers worldwide, successfully automating video integration across hundreds of thousands of web pages. To date, the Contextual Video Matching has generated billions of video views, proving its ability to maintain high engagement at an industrial scale.

This broad adoption translates into significant business value. By enabling high-quality video placements on pages that were previously not monetized by Dailymotion, Contextual Video Matching now accounts for 2\% to 3\% of Dailymotion’s revenue coming from publishers sites. This demonstrates that LLM-grounded contextual matching is not only a technical improvement but a primary driver of our global ad-driven revenue streams.

\section{Conclusion \& Future work}

In this paper, we presented Contextual Video Matching, an automated system designed to bridge the semantic gap between full-length editorial articles and video content, a real-world issue for our publishers who produce a large volume of both videos and web pages. This system successfully increases user engagement and enriches the overall user experience, while providing clear value to the publishers. For Dailymotion, it also represents a new source of revenues as well as a way to extend the number of places Dailymotion player is used, improving our brand visibility. 

While already very successful, the Contextual Video Matching system can be improved in a number of ways and a lot of work remains as this is only an initial step toward matching videos with long-form text. For instance, the system currently relies solely on the textual information and could benefit from leveraging multi-modal data \cite{liu2024multimodal, denadai2025describe}, especially on the video side, for which we could consider using audio or visual features from video thumbnails and extracted frames. In this direction, the next step for us will likely be to improve the representativeness of the videos' embeddings by using not only their textual metadata, but also their transcript which we already compute using a speech-to-text model, Whisper \cite{radford2023whisper}. Finally, future research might explore learnable ranking functions to further refine the alignment between editorial intent and video suggestions.

\bibliographystyle{ACM-Reference-Format}
\bibliography{references} 

\end{document}